\documentclass[12pt]{article}
\usepackage{graphicx,amsmath}
\usepackage{units}
\usepackage{color}

\newlength{\dinwidth}
\newlength{\dinmargin}
\def\be{\begin{equation}}   
\def\ee{\end{equation}}  
\def\bea{\begin{eqnarray}}                      
\def\eea{\end{eqnarray}}
\def\ch1{$\chi(1^+)$}
\def\lapproxeq{\lower .7ex\hbox{$\;\stackrel{\textstyle                                                    
<}{\sim}\;$}}                                                    
\def\gapproxeq{\lower .7ex\hbox{$\;\stackrel{\textstyle                                                    
>}{\sim}\;$}}

\begin{document}

\begin{flushright}                                                    
IPPP/16/07  \\
DCPT/16/14 \\                                                    
\today \\                                                    
\end{flushright} 

\vspace*{0.5cm}

\begin{center}

{\Large \bf Advantages of exclusive $\gamma\gamma$ production to probe} \\
\vspace{0.5cm}
{\Large \bf high mass systems}

\vspace*{1cm}

A.D. Martin$^a$ and M.G. Ryskin$^{a,b}$ \\

\vspace*{0.5cm}

$^a$ Institute for Particle Physics Phenomenology, University of Durham, Durham, DH1 3LE \\
$^b$ Petersburg Nuclear Physics Institute, NRC `Kurchatov Institute', Gatchina, St.~Petersburg, 188300, Russia \\

\begin{abstract}
We recall that the {\it exclusive} production of high mass objects via $\gamma\gamma$ fusion at the LHC is not strongly suppressed in comparison with {\it inclusive} $\gamma\gamma$ fusion.  Therefore it may be promising to study new objects, $X$, produced by the $\gamma\gamma$ subprocess in experiments with exclusive kinematics.  We list the main advantages of exclusive experiments. We discuss the special advantage of observing $\gamma\gamma \to X \to \gamma Z$ exclusive events.
\end{abstract}

\end{center}
\vspace{0.5cm}

\section{Introduction}
Observation of a possible peak in the $\gamma\gamma$ mass distribution at 750 GeV \cite{A,C} has focused interest on processes driven by $\gamma\gamma$ fusion.  Indeed a number of Authors recently argued that $\gamma\gamma$ fusion may dominate the production of new diphoton resonances \cite{Royon,Tern,Fich}, see also \cite{LAA}$-$\cite{d}.   The present data are for {\it inclusive} kinematics, so the $\gamma\gamma$ system will be accompanied by secondaries produced in the same process and by the underlying event.  It was emphasized \cite{Royon,Fich} that a cleaner kinematic environment may be provided by an {\it exclusive} experiment in which the leading protons are tagged and only the resonance or some low multiplicity system $X$ is produced in the $\gamma\gamma\to X$ subprocess.  That is, we deal with the process $pp\to p+X+p$ where the + signs  denote large rapidity gaps.  In this short note, we describe the most attractive features that are provided by exclusive kinematics.

\section{Exclusive $\gamma\gamma$ production}

Exclusive production has been discussed in many papers, see, for example, \cite{Pros}$-$\cite{s}, but mainly for subprocesses driven by gluon-gluon fusion.  Here the problem is that the effective $gg$ luminosity, ${\cal L}_{gg}$, decreases steeply with $M_X$, ${\cal L}_{gg}\sim 1/M_X^{3.3}$. Thus the cross section for the exclusive production of a heavy system $X$ becomes too small.  This is due to the presence of the Sudakov factor $T$ -- that is, the probability that there is no additional gluon bremsstrahlung in the fusion of the high-mass $gg$ pair into a colourless object.  The higher the mass, the greater the phase space for gluon bremsstrahlung, the smaller the exclusive cross section. 

Exclusive $\gamma\gamma$ fusion does not suffer from this problem.  Indeed, it was shown in \cite{Pros} that already for $M_X>200$ GeV, the effective $\gamma\gamma$ flux exceeds that for gluons. More precise calculations of the $\gamma\gamma$ flux have recently been performed \cite{HKR}.  Since the $\gamma$-exchange amplitude arises mainly from large impact parameters where the opacity of the proton is already small, there is almost no suppression, $S^2$,  due to soft proton rescattering which could otherwise have produced additional secondaries in the underlying event. Moreover, there is no QCD Sudakov $T$ factor in colourless $\gamma\gamma$ fusion.  Thus in comparison to inclusive production, the exclusive rate for $\gamma\gamma$ fusion  is only suppressed by a factor of 14 for $M_X\sim 750$ GeV \cite{HKRdiphoton}.   (Note that the corresponding suppression is more than 5 orders of magnitude for gluon-gluon fusion.)   Therefore it is not unrealistic to study such a heavy object in an exclusive $\gamma\gamma$ fusion experiment.

For high luminosity at the LHC there will be up to 50$\sim$100 pile-up events per bunch crossing.  Nevertheless it should be possible to select an exclusive event by matching the momenta measured in the forward proton detectors with the whole momentum of the system $X$ observed in the central detector.  Moreover, good timing of the forward proton detectors will allow the $z$ position (that is, the position along the beam line) of the vertex to be fixed, which should coincide with that of the centrally produced particles.\footnote{Unfortunately the high $E_T$  photons are detected by the
calorimeter and therefore the $z$-position of the vertex from the
central detector will be known with low accuracy.
Moreover, the signal may be smeared by other particles
which hit the same calorimeter cell.
However, the last problem does not cause a difficulty for 
photon fusion induced events. In such a case the probability that
the hard process is accompanied by additional jet activity is quite
small (see \cite{HKRdiphoton}) while the secondaries from the underlying
events have low $p_t$ and a flat rapidity distribution in the central
region. Finally, in sect. \ref{sec:other}, we discuss another decay channel (in
particular, $X\to \gamma Z$) where the position of the vertex can
be seen from the charged tracks in the central detector from this decay.}

The next question is the acceptance of the detectors, especially those registering the forward protons. Taking $M_X\sim 750$ GeV as an example, we see the momentum fraction lost by the proton is
\be
\xi~=~(M_X/\sqrt{s})~e^{y}~\simeq 0.06~e^{y},
\ee
for the LHC at 13 TeV.
Thus for events with the $X$ system lying in the rapidity interval $-2<y<2$ we need to have good efficiency of the forward detectors in the interval $0.01\lapproxeq \xi \lapproxeq 0.4$.   These events will be mainly concentrated in the proton transverse momentum interval $50\lapproxeq p_t\lapproxeq 700$ MeV.  Such large intervals are not completely covered by the present forward detectors \cite{Yellow}.   Realistically we may hope to  collect only about a third of the exclusive events.  Thus assuming an inelastic cross section of diphoton production from $\gamma\gamma$ fusion of 6 fb, and an eventual integrated LHC luminosity of 300 fb$^{-1}$, we will detect about $300\times 6/(14\times 3)\simeq 40$ exclusive events.

\section{Advantages of exclusive events}
The main advantages of performing experiments which collect exclusive $\gamma\gamma$ events are:

\subsection{Negligible background}
In spite of the fact that the cross section for the exclusive $\gamma\gamma$-fusion  process, $pp \to p+X+p$ is about an order of magnitude smaller than the signal with inclusive kinematics, the exclusive signal will be more visible due to the very strong suppression of the background. In particular in \cite{Pros} it was found the $\gamma\gamma$ luminosity already exceeds the exclusive $gg$ luminosity for $M_X\gapproxeq 200$ GeV. As mentioned above, the reason is that the exclusive $gg$ luminosity is strongly suppressed by the Sudakov factor -- the factor for preventing additional gluon bremsstrahlung and responsible for the behaviour ${\cal L}_{gg}\sim 1/M_X^{3.3}$. For $M_X \sim 750$ GeV the $gg$ luminosity is about two orders of magnitude smaller than the $\gamma\gamma$ luminosity \cite{Pros, HKRdiphoton}.
That is, even with the hard subprocess enhanced by the coupling factor $(\alpha_s/\alpha_{\rm QED})^2 \sim  100$, the photon induced exclusive cross section may dominate.

In particular, the QCD induced exclusive $\gamma\gamma$ background will be less than $3\times 10^{-4}$ fb already at $M_X=110$ GeV \cite{KMRS}. In other words the exclusive $\gamma\gamma$-fusion signal has practically no QCD background!

\subsection{Mass resolution}
An exclusive process has the unique advantage that the mass of the system $X$ can be measured by two independent methods.  In addition to the mass, $M_X$, of the observed final state system, we now have the possibility to also calculate $M_X$ using the {\it missing mass} from the momenta of the protons observed in the forward detectors. This provides an important experimental constraint. It will improve the mass resolution.  Moreover for a wide peak, composed of two or more narrow resonances, there may be the possibility to unfold these states (see, for example, \cite{KKMR1,Hein})

\subsection{Other decay channels  \label{sec:other}}
A strong suppression of the background provides a good environment to study resonance production in other decay channels. An attractive channel to supplement the $\gamma\gamma$ decay, is $X\to \gamma Z$ decay \cite{Royon,new}.  The $Z$ boson can be observed in an exclusive experiment, not only by its leptonic decay channels, but also via the $Z\to q\bar{q}$ decay. That is, it may be possible to collect almost all of the $Z$ signal by looking for events where the transverse momentum of the decay $\gamma$ is balanced by two high $p_t$ jets. The two $Z$ decay jets will not be collinear, but rather separated by an angle $\theta \sim 2M_Z/M_X$.

Indeed, the observation of the $X \to \gamma Z$ decay has an important advantage.  Modern precise timing of the forward proton detectors should allow a determination of the interaction vertex along the beam line with a resolution of about a mm or so. This must coincide with the {\it charged tracks} observed in the central detectors from $Z$ decay.  So this decay channel provides a powerful way to avoid pile-up problems and to identify clean exclusive events.

\subsection{Spin-parity analysis}
As was discussed in \cite{KKMR,HKRdiphoton} the azimuthal correlations between the transverse momenta of the forward protons greatly improve the possibility of a spin-parity analysis of any exclusively produced heavy resonance.  In particular, recall that it is impossible to distinguish between scalar and pseudoscalar particles simply studying the distributions of their decay products.  In both cases the distribution is isotropic. On the other hand, the azimuthal correlation of the forward proton momenta will have a maximum at $\phi=0$ (or $\pi$) for a scalar state, but at $\phi=\pi/2$ (or $3\pi/2$) for a pseudoscalar state.

Moreover, it was shown in \cite{KMRcp} that an asymmetry  in the forward proton distributions should be observed if there is CP-violation in the coupling of the heavy state to $\gamma\gamma$.

\section*{Acknowledgments}

We thank Albert de Roeck for valuable comments.   MGR thanks the IPPP at the University of Durham for hospitality. This work 
was supported by the RSCF grant 14-22-00281.

\end{document}